# IDENTIFICATION AND CHARACTERIZATION OF TWO ENERGY GAPS IN SUPERCONDUCTING MgB$_2$ BY SPECIFIC–HEAT MEASUREMENTS

R. A. Fisher,[1] F. Bouquet,[1] N. E. Phillips,[1] D. G. Hinks,[2] and J. D. Jorgensen[2]

[1]Lawrence Berkeley National Laboratory and Department of Chemistry, University of California, Berkeley, CA 94720 USA
[2]Materials Science Division, Argonne National Laboratory, Argonne, IL 60439 USA

## 1. INTRODUCTION

In the BCS theory of superconductivity [1] the electron pairing is mediated by phonons, and high phonon frequencies favor high values of the critical temperature ($T_c$). The small atomic mass of boron suggests the likelihood of high vibrational frequencies in its compounds. Although a number of boron intermetallic compounds were investigated for superconductivity in the 1970's, the observed $T_c$'s were unremarkable [2]. The synthesis of MgB$_2$ had been reported in 1954 [3] but evidently it was not included in those investigations. It is now available commercially in substantial quantities, and used as a metathesis reagent in the synthesis of other borides. In light of this history, the seemingly accidental [4] discovery [5] of superconductivity in MgB$_2$, and particularly the high value of $T_c$, ~ 39 K, was a great surprise. The isotope effect [6–8] suggests that phonon–driven coupling of the electrons is important, but the magnitude of $T_c$ seems to be above the limit that is expected for that mechanism [9]. Measurements of the specific heat ($C$) can give information relevant to understanding the mechanism: They give some information on the phonon spectrum, but the details are obscured by the averaging in $C$; more importantly, they give the electron density of states (EDOS), and comparison with band–structure calculations gives the electron–phonon coupling parameter ($\lambda$). Both of these parameters are important in determining $T_c$ in phonon–mediated superconductors. For MgB$_2$ their values would be consistent with a purely phonon–mediated mechanism only if there were some other unusual factor at work [10]. Specific–heat data do show unambiguously the existence of one unusual feature, a low–temperature exponential dependence on temperature that shows the presence of a second energy gap that is ~ 1/4 the magnitude of that expected for the value of $T_c$ [10]. Other specific-heat measurements have given similar results [11, 12] confirming that this feature is intrinsic. Energy gaps in superconductors are usually investigated by spectroscopic techniques, which are subject to errors associated with surface effects, but in MgB$_2$ the gap structure is so pronounced that it can be characterized by measurements of the specific heat, a bulk property. Although two–gap superconductivity has been recognized as a theoretical possibility and some of its manifestations predicted [13–15], these measurements were the most convincing experimental evidence of its existence. Subsequent spectroscopic studies have, in some cases, also shown the presence of two gaps (see Sec. 5, below). The different magnitudes of the two gaps are indicative of substantially different electron–phonon coupling strengths on two parts of the Fermi surface, which, together with the inter–band couplings, determine $T_c$. Recent calculations of the phonon spectrum and electron–phonon coupling parameters suggest that the high value of $T_c$ is a consequence of the contribution to the electron-phonon coupling from scattering between two pairs of sheets of the Fermi surface [16]. The calculations give two gaps, characterized by parameters that are in excellent agreement with those deduced from specific–heat data (see below).

## 2. SAMPLE AND SPECIFIC–HEAT MEASUREMENTS

The Mg$^{11}$B$_2$ sample was a powder, prepared by reacting $^{11}$B powder and Mg metal in a capped BN crucible at 850ºC under a 50–bar argon atmosphere for 1.5 hours. Thermal contact to the powder was achieved by mixing it with a small amount of GE7031 varnish (a common low–$T$, thermal–contact agent with a known heat capacity) in a thin–walled copper cup. These extra contributions to the addenda limited the precision of the data by reducing the contribution of the sample to between 10 to 75% of the total measured heat capacity. However, the alternate method of providing thermal contact, sintering the powder, can have adverse effects on the sample [17] and may in some cases account for differences between the results reported here and those obtained in other measurements. The measurements of $C$ were made by a modified heat–pulse technique, 1–32 K, and by a continuous–heating technique, 29–50 K. Varying the heating rate by 50% had no effect on the continuous–heating results, verifying the absence of errors associated with thermal–relaxation times. Measurements were made in magnetic fields ($H$), to 9 T, on the field–cooled (FC) sample after applying the fields at $T \geq 60$ K. For $H = 1$ T, for which the field penetration is about 50% of that in the normal state, measurements were also made after applying the field at 1 K. The results were indistinguishable from the FC results and there were no indications of irreversibility, suggesting that equilibrium flux distributions were attained. This behavior might be a property of the powder, as distinct from the behavior of sintered samples.

## 3. ANALYSIS OF SPECIFIC–HEAT DATA

The specific heat is the sum of an $H$–dependent electron contribution ($C_e$) and an $H$–independent lattice contribution ($C_l$). In the normal state, $C_e(H) = \gamma_n T$, independent of $H$; in the mixed state, $C_e(H)$ includes a $T$–proportional term, $\gamma(H)T$, and other $H$–dependent terms; in the superconducting state, $C_e(0) = C_{es}$. Most of the interpretation of the results is based on an analysis of $C(H)$ - $C(9$ T). This has the advantage that $C_l$ and most of the heat capacity of the addenda, which is independent of H, cancel. This leaves $C_e(H)$ - $C_e(9$ T), and it is $C_e$ that is the quantity of most interest. Furthermore, $C_e(9$ T$) \approx \gamma(9$ T$)T$ (see below and Ref. 10), $C(H)$ - $C(9$ T$) \approx C_e(H)$ - $\gamma(9$ T$)T$, and the quantity plotted in Fig. 1, $[C(H)$ - $C(9$ T$)]/T + \gamma(9$ T$)$, is approximately equal to $C_e(H)/T$.

The upper critical field ($H_{c2}$) of MgB$_2$ is approximately linear in $T$ with $H_{c2}(0) \sim 16$ T [18]. This is reasonably consistent with the measurements of C, for which the onset of the transition to the mixed state, at $H = H_{c2}(T)$, is marked by the deviations of $C(H)$ - $C(9$ T) from zero (see Fig. 1(a)). It leads to the expectation that for 9 T the sample would be in the normal state for $T \geq 20$ K. The data actually plotted in Fig. 1(a) are $[C(H)$ - $C(9$ T$)]/T$, but with the scale shifted by $\gamma(9$ T). (They are essentially point–to–point differences, with no smoothing of the 9–T data, which increases the scatter.) The values of $\gamma(H)$ have been determined by fitting the low–$T$, mixed– and superconducting–state data with $[C(H)$ - $C(9$ T$)] = [\gamma(H)$ - $\gamma(9$ T$)]T + a\exp(-b/T)$, where $a$ and $b$ are $H$ dependent. Consistent with the $H = 0$ data, $\gamma(0)$ was taken as zero, fixing the values of $\gamma(H)$ for all $H$. Figure 2 shows $\gamma(H)$ $vs.$ $H$, and the extrapolation to $H_{c2}(0) = 16$ T to obtain $\gamma_n = 2.6$ mJ mol$^{-1}$ K$^{-2}$. Although the extrapolation is somewhat arbitrary, the very small differences between $\gamma(5$ T), $\gamma(7$ T), and $\gamma(9$ T) suggest that it gives $\gamma_n$ to within $\sim 0.1$ mJ mol$^{-1}$ K$^{-2}$. If the sample were normal at all temperatures in 9 T the quantity shown in Fig. 1(a) would be exactly $C_e(H)/T$. That is not the case, but the differences are small, as shown by the small differences between $C(5$ T), $C(7$ T), and $C(9$ T), and as described in detail in Ref. 10.

The transition to the superconducting state, shown in Fig. 1(c), with an entropy–conserving construction that gives $T_c = 38.7$ K and $\Delta C(T_c) = 133$ mJ mol$^{-1}$ K$^{-1}$, is relatively sharp, with a width $\sim 2$ K. In this temperature interval the sample is in the normal state for $H = 9$ T, and addition of $\gamma_n = 2.6$ mJ mol$^{-1}$ K$^{-2}$ to the quantity plotted would give $C_e(H)$ through the transition. The effect of $H$ in broadening the transition (to the mixed state), as expected for measurements on a powder with an anisotropic $H_{c2}$, is evident in Fig. 1(a).

The thermodynamic consistency of the data, including in particular the very unusual $T$ dependence of $C_{es}(0)$, can be tested by calculating the difference in entropy ($S$) between 0 and $T_c$ for different fields. The result of such a test is shown in Fig. 3(a) where the entropies obtained by integrating the plotted points are compared with $\gamma$(9 T)$T$, which represents the 9–T data. At 40 K, the entropies for all $H$ are within ± 2% of the same value. The result of a second integration of the entropies to obtain free energies and the thermodynamic critical field ($H_c$) is shown in Fig. 3(b).

Fitting the 9–T data for $20 \leq T \leq 50$ K with $C(9 \text{ T}) = \gamma_n T + C_l$, where $C_l = B_3 T^3 + B_5 T^5$, gave $B_3 = 5.1 \times 10^{-3}$ mJ mol$^{-1}$ K$^{-4}$ and $B_5 = 2.5 \times 10^{-6}$ mJ mol$^{-1}$ K$^{-6}$. The Debye temperature ($\Theta$), calculated following the usual convention of using the value of $B_3$ per g atom, $\Theta^3 = (12/5)\pi^4$ R $(3/B_3)$, is $1050 \pm 50$ K.

## 4. DISCUSSION

Anisotropy in $H_{c2}$ cannot explain the dramatic increase in $\gamma(H)$ at low $H$ shown in Fig. 2. The dashed curve is a calculation using the effective–mass model with an anisotropy of 10, which is already greater than reported values [19, 20], but $\gamma(H)$ cannot be fitted with *any* value of the anisotropy. The rapid increase in $\gamma(H)$ at low $H$ is a consequence of the small magnitude of the second gap and the associated small condensation energy.

An average over the Fermi surface of the electron–phonon coupling parameter can be estimated by comparing $\gamma_n$ with band–structure calculations of the "bare" EDOS at the Fermi surface ($N(0)$) using the relation $\gamma_n = (1/3) \pi^2 k_B^2 N(0)(1+\lambda)$. $N(0)$ has been reported as 0.71, 0.68, and 0.72 states eV$^{-1}$ unit cell$^{-1}$ (Refs. 19, 21, 22, respectively) giving $\lambda = 0.51$, 0.58, and 0.49. Theoretically calculated values of $\lambda$ are 0.71 [21], and ~ 1 [22]. $T_c$ is related to $\lambda$, $\Theta$, and the electron–electron repulsion ($\mu^*$) by [9]

$$T_c = (\Theta/1.45) \exp\{-1.04(1 + \lambda)/[\lambda - \mu^*(1 + 0.62\lambda)]\}, \qquad (1)$$

with $\mu^*$ frequently taken to be ~ 0.1 [9, 23]. With $\lambda = 0.58$, the highest of the values obtained from $\gamma_n$ and $N(0)$, and $\Theta = 1050$ K, Eq. (1) gives $T_c = 18$ K, too low by more than a factor two. For these values of $\lambda$ and $\Theta$, $T_c = 39$ K would require $\mu^* = 0.013$, an unusually low value. However, in this expression $\Theta$ represents a relatively crude estimate of the phonon frequencies that are important in the electron pairing. As derived from the coefficient of the $T^3$ term in $C_l$, $\Theta$ is really a measure of the frequencies of the low–frequency acoustic phonons, which may not be particularly relevant to the pairing of the electrons. In the Debye model, $\Theta$ is also the cut–off frequency, but real phonon spectra often extend to significantly higher frequencies. In another version of the theory [23], $\Theta/1.45$ is replaced by $\omega_{\log}/1.20$, where $\omega_{\log}$ is a moment of the phonon frequencies in which they are weighted by the electron–phonon matrix elements. More detailed calculations [16] that take into account relevant features of the phonon spectrum and electron–phonon scattering account for the observed $T_c$ with a physically plausible value of $\mu^*$.

The parameters $[H_c(0)]^2/\gamma_n T_c^2$ and $\Delta C(T_c)/\gamma_n T_c$ are correlated with the strength of the electron pairing [24]. In the BCS, weak–coupling limit, their values are 5.95 and 1.43, respectively. There are a number of "strong–coupled" superconductors for which these parameters are greater than the BCS values, but relatively few for which they are smaller [24]. For Mg$^{11}$B$_2$ they are unusually small, 5.46 and 1.32. Such small values can arise from gap anisotropy (see *e.g.*, Ref. 24) and, at least in the case of $\Delta C(T_c)/\gamma_n T_c$, from multi–gap structure [13].

Figure 1(a) includes a comparison of the experimental $C_{es}$ with that for a BCS superconductor with $\gamma_n = 2.6$ mJ mol$^{-1}$ K$^{-2}$ and $T_c = 38.7$ K. For $T \geq 27$ K, $C_{es}$ is approximately parallel to the BCS curve; at lower temperatures it rises above the BCS curve and then decreases to zero as $a$exp($-b/T$), but with values of $a$, 67 mJ mol$^{-1}$ K$^{-1}$, and $b$, 15.8 K, very different from those of the BCS superconductor. This behavior is conspicuously different from that of any other superconductor. It gives the appearance of a transition to the superconducting state in two stages, associated with two energy gaps: the first, a partial transition that leaves a "residual" $\gamma$ (the extrapolation of $C_{es}$ to $T = 0$ from above 12 K gives ~ 1 mJ mol$^{-1}$ K$^{-2}$); the second, which completes the transition and is associated with a second, smaller energy gap. The low–$T$ exponential temperature dependence is shown more clearly in Fig. 4. $C_{es}$ is well represented by a simple exponential over a much wider range of temperatures, $4 < T_c/T < 17$, than for a BCS superconductor, which suggests a weaker–than–BCS temperature dependence of the small gap. A comparison of the value of $b$ with BCS expressions valid for the temperature interval in which $b$ was determined gives, as a rough approximation to the $T = 0$ gap parameter for the small gap, $\Delta_2(0) = 0.44$k$_B T_c$, about one fourth of the BCS value. For a two–band, two–gap superconductor, interband coupling ensures that the two gaps open at the same $T_c$ [13]. If the electron–phonon coupling is weaker in one band than the other, as suggested in this case by the ratio of the gaps, the two gaps are likely to have similar temperature dependences [15] but $C_{es}$ will be determined by the smaller gap at low temperature [13]. At least qualitatively the temperature dependence of $C_{es}$ is consistent with expectations for a two–gap superconductor with the two gaps differing in amplitude by a factor ~ 4 (but see below for a different estimate).

A model for a two–gap superconductor [25], which is a generalization of a successful semi–empirical model for strong–coupled, single–gap superconductors [24], provides the basis for a more quantitative interpretation of $C_{es}$. The general expression [13] for the superconducting–state entropy of two–band, two–gap superconductors, in the notation of Ref. 24, is

$$S_{es} = -2\text{k}_B \, 3 \, 3 \, [f_k \ln f_k + (1 - f_k)\ln(1 - f_k)], \qquad (2)$$

where the double sum is over the quasiparticle states in both bands, $f_k = [\exp(E_k/\text{k}_B T) + 1]^{-1}$, and the $E_k$ are the quasiparticle energies, which are different in the two bands. $E_{k1}^2 = \varepsilon_{k1}^2 + \Delta_1^2$ and $E_{k2}^2 = \varepsilon_{k2}^2 + \Delta_2^2$, where the $\varepsilon_k$ are the normal–state quasiparticle energies, and $\Delta_1(T)$ and $\Delta_2(T)$ are the gap functions. Using Eq. (2), Bouquet *et al.* [25] generalized the "$\alpha$ model" [24] to two gaps, approximating the temperature dependence of both gaps with that of a weak–coupled BCS superconductor. The gaps scale with the BCS gap, $\Delta_1(T) = (\alpha_1/\alpha_{BCS})\Delta_{BCS}(T)$ and $\Delta_2(T) = (\alpha_2/\alpha_{BCS})\Delta_{BCS}$, where $\alpha_{BCS} = 1.764$ and $\alpha_1$ and $\alpha_2$ are adjustable parameters. The reduced gaps, $\delta(T) = \Delta_1(T)/\Delta_1(0) = \Delta_2(T)/\Delta_2(0)$, have the numerical values tabulated by Mhhlshlegel [26]. Equation (2) of Ref. 24 then becomes, for a two–gap superconductor,

$$S_{es}(t)/\gamma_n T_c = - (3/\pi^2)\alpha_1(\gamma_1/\gamma_n) \! \int \! dx[f_{x1}\ln f_{x1} + (1 - f_{x1})\ln(1 - f_{x1})]$$
$$- (3/\pi^2)\alpha_2(\gamma_2/\gamma_n) \! \int \! dx[f_{x2}\ln f_{x2} + (1 - f_{x2})\ln(1 - f_{x2})] \qquad (3)$$

where $\gamma_1$ and $\gamma_2$ represent the EDOS in the two bands, $f_{x1} = \{\exp[\alpha_1 t^{-1}(x^2 + \delta^2)^{1/2}] + 1\}^{-1}$, $f_{x2} = \{\exp[\alpha_2 t^{-1}(x^2 + \delta^2)^{1/2}] + 1\}^{-1}$, and $t \equiv T/T_c$. There are four new parameters, $\alpha_1$, $\alpha_2$, $\gamma_1$, and $\gamma_2$, but one constraint, imposed by the total (experimentally–determined) normal–state EDOS, $\gamma_1 + \gamma_2 = \gamma_n$.

Equation (3) was integrated numerically for combinations of values of $\alpha_1$, $\alpha_2$, $\gamma_1$, and $\gamma_2$, and $C_{es}$ was then obtained by numerical differentiation [25]. The "best fit", obtained for $\alpha_1 = 2.2$, $\alpha_2 = 0.60$, $\gamma_1 = 0.55\gamma_n$ and $\gamma_2 = 0.45\gamma_n$, is compared with the experimental data in Fig. 5. It corresponds to 55% of the total EDOS associated with one sheet of the Fermi surface, for which the gap parameter is $\Delta_1(0) = 1.2\Delta_{BCS}(0) = 7.3$ meV, and 45% of the EDOS associated with a second sheet of the Fermi surface, for which the gap parameter is $\Delta_2(0) = 0.34\Delta_{BCS}(0) = 2.0$ meV. These results are consistent with several constraints imposed by the general theory of two–gap superconductors: in the low–temperature limit, $C_{es}$ is determined by the small gap [13] (the simple exponential fit to those data gives $\Delta_2(0) = 0.25\Delta_{BCS}(0)$); $\Delta C(T_c)/\gamma_n T_c$ must be less than the BCS value [13] (the experimental value is 1.32 and the BCS value is 1.43); one gap must be larger than the BCS gap and one smaller [15] ($\Delta_1(0) = 1.2\Delta_{BCS}(0)$ and $\Delta_2(0) = 0.34\Delta_{BCS}(0)$). The ratio $\Delta_1(0)/\Delta_{BCS}(0) = 1.2$ can be a consequence of the interband coupling, and does not necessarily imply strong coupling in that band. The derived values of the four parameters are in remarkably good agreement with the band–structure calculations of Liu *et al.* [16] who find the EDOS distributed 53% on one sheet of the Fermi surface with a gap parameter $2.0 k_B T_c$ and 47% on another with a gap parameter of $0.65$ $k_B T_c$. The same calculation [16] shows deviations from BCS temperature dependence for each gap, which can be expected as a consequence of the interband coupling. These deviations, and those observed in spectroscopic determinations of the temperature dependence of the gap (see below), are relatively small in comparison with the differences in amplitudes of the two gaps, and would probably have only a small effect on $C_{es}$. The agreement of the specific–heat data with the two–gap model and the agreement of the derived parameters with theoretical calculations argue persuasively for both the existence of two–gap superconductivity in $MgB_2$ and its relation to the high value of $T_c$.

## 5. SPECTROSCOPIC DETERMINATIONS OF THE ENERGY GAPS

The majority of the spectroscopic determinations of the gap parameters for $MgB_2$ identify only one gap, but a significant number do show two gaps, which in a few cases are in approximate agreement with the $\Delta_1(0) = 7.3$ meV and $\Delta_2(0) = 2.0$ meV obtained by fitting $C_{es}$. Scanning tunneling [27–32], point-contact [33, 34], Raman [35], and photoemission [36, 37] spectroscopies have all been used. *Scanning tunneling spectroscopy*: Measurements made at liquid–helium temperatures have shown gaps ranging from $\Delta(0) = 2.0$ to 7.8 meV. These gaps tend to be V–shaped for sintered samples [27–29], but "flat bottomed" for films, powders, or small crystals [30–32]. Etching the samples reduces the scatter in the data (presumably by removing surface contamination), and for sintered samples the V–shape evolves towards a flat bottom [29]. One set of measurements [31] on a powder resolved two gaps, $\Delta_1(0) = 7.5$ meV and $\Delta_2(0) = 3.9$ meV. For a small single crystal [32] that showed two gaps, $\Delta_1(0) = 7.5$ meV and $\Delta_2(0) = 3.5$ meV, the temperature dependence of each gap, measured from 4 to 35 K, was BCS-like. *Point–contact spectroscopy*: One measurement [33] showed a single gap with $\Delta(0) = 4.3$ to 4.6 meV, while in another experiment [34], on a granular powder, two gaps were observed with $\Delta_1(0) = 7.0$ meV and $\Delta_2(0) = 2.8$ meV. Their temperature dependence, measured from 4 to 40 K, was BCS–like. *Raman spectroscopy*: A powdered sample showed two gaps, $\Delta_1(0) = 6.2$ meV and $\Delta_2(0) = 2.7$ meV [35]. *Photoemission spectroscopy*: One high-resolution study [36] gave only a single gap (perhaps because the low-$T$ measurements did not extend below 15 K), $\Delta(0) = 4.5$ meV at 15 K, with a temperature dependence between 15 and 40 K that followed the BCS form. Another experiment [37] resolved two gaps with $\Delta_1(0) = 5.6$ meV and $\Delta_2(0) = 1.7$ meV and an essentially BCS temperature dependence. Quite generally then, the temperature dependence of the gaps, when observed, is, at least approximately, BCS–like, as was assumed in the two–gap model for $C_{es}$.

## 6. OTHER SPECIFIC–HEAT MEASUREMENTS

Most other measurements of the specific heat do not give evidence that bears on the existence of a second energy gap, but two of them [11, 12] do show behavior of $C_{es}$ in the region below 10 K that is similar to that shown in Fig. (1a). In one case [12] essentially the same exponential temperature dependence was observed; in the other [11] the temperature dependence of $C_{es}$ was obscured by impurity contributions below $\sim 4$ K and the low–$T$ temperature dependence was taken as approximately $T^2$. These differences notwithstanding, all three of these measurements give evidence of the same small gap, with $\Delta_2(0) \sim 2.0$ meV. All three of the measurements give essentially the same value of $\gamma_n$, 2.7 mJ mol$^{-1}$ K$^{-1}$ [11, 12] or 2.6 mJ mol$^{-1}$ K$^{-1}$, reported here and in Ref. 10. In each case the value is supported by tests of the thermodynamic consistency of the normal– and superconducting–state specific heats.

Several other values of $\Delta C(T_c)$, all for sintered samples and lower than that reported here and in Ref. 10, 133 mJ mol$^{-1}$ K$^{-1}$, have been given in other reports: 119 mJ mol$^{-1}$ K$^{-1}$ [6, 12]; 77 mJ mol$^{-1}$ K$^{-1}$ [11]; 76 mJ mol$^{-1}$ K$^{-1}$ [38]. One higher value, $\sim 140$ mJ mol$^{-1}$ K$^{-1}$, on a sintered sample, has also been reported [39] but it is less well defined by the experimental data than most of the others. Values of $\gamma_n$ that range from 1.1 (but that value, which has been quoted elsewhere, has been revised to 3.1 mJ mol$^{-1}$ K$^{-2}$ [38]) to 5.5 mJ mol$^{-1}$ K$^{-2}$, based on different [38, 39] or unspecified [6] analysis of experimental data have also been reported, but the correct value would seem to be 2.6–2.7 mJ mol$^{-1}$ K$^{-2}$.

A more comprehensive review of specific–heat measurements is also given in this volume [40].

## ACKNOWLEDGEMENTS

We have benefited from useful discussions with J. M. An, M. L. Cohen, G. W. Crabtree, J. P. Franck, R. A. Klemm, V. Z. Kresin, S. G. Louie, C. Marcenat, I. I. Mazin, and W. E. Pickett, and D. Roundy. We have profited particularly from discussions on two–band superconductors with V. Z. Kresin. The work at LBNL was supported by the Director, Office of Basic Energy Sciences, Materials Sciences Division of the U. S. DOE under Contract No. DE-AC03-76SF00098. The work at ANL was supported by the U. S. DOE, BS-MS under Contract No. W-31-109-ENG-38.

## REFERENCES

[1] J. Bardeen, L. N. Cooper, and J. R. Schrieffer, Phys. Rev. **108** (1957) 1175.

[2] A. S. Cooper, E. Corenzwit, L. D. Longinotti, B. T. Matthias, and W. H. Zachariasen, Proc. Natl. Acad. Sci. USA **67** (1970) 313; L. Leyarovska, and E. A. Leyarovski, J. Less Common Metals **67** (1979) 249.

[3] M. Jones and R. Marsh, J. Am. Chem. Soc. **76** (1954) 1434.

[4] C. Day, Physics Today **54** (2001) 17.

[5] J. Nagamatsu, N. Nakagawa, T. Muranaka, Y. Zenitani, and J. Akimitsu, Nature **410** (2001) 63.

[6] S. L. Bud'ko, G. Lapertot, C. Petrovic, C. E. Cunningham, N. Anderson, and P. C. Canfield, Phys. Rev. Lett. **86** (2001) 1877.

[7] D. G. Hinks, H. Claus, and J. D. Jorgensen, Nature **411** (2001) 457.

[8] D. D. Lawrie, J. P. Franck, and G. Zhang to be published.

[9] W. L. McMillan, Phys. Rev. **167** (1968) 331.

[10] F. Bouquet, R. A. Fisher, N. E. Phillips, D. G. Hinks, and J. D. Jorgensen, Phys. Rev. Lett., scheduled for the July 23, 2001 issue.

[11] Y. Wang, T. Plackowski, and A. Junod, Physica C **355** (2001) 179.

[12]     H. D. Yang, J.–Y. Lin, H. H. Li, F. H. Hsu, C. J. Liu, and C. Jin, cond–mat/0104574 (2001).

[13]     B.T. Geilikman, R. O. Zaitsev, and V. Z. Kresin, Sov. Phys. Solid State **9** (1967) 642.

[14]     V. Z. Kresin, J. Low Temp. Phys. **11** (1973) 519.

[15]     V. Z. Kresin and S. A. Wolf, Physica C **169** (1990) 476.

[16]     A. Y. Liu, I. I. Mazin, and J. Kortus, cond–mat/0103570 (2001).

[17]     J. P. Franck, private communication.

[18]     S. L. Bud'ko, C. Petrovic, G. Lapertot, C. E. Cunningham, P. C. Canfield, M.– H. Jung, and A. H. Lacerda, Phys. Rev. B **63** (2001) 220503R.

[19]     J. M. An and W. E. Pickett, Phys. Rev. Lett. **86** (2001) 4366.

[20]     O. F. de Lima, R. A. Ribeiro, M. A. Avila, C. A. Cardosa, and A. A. Coelho, cond–mat/0103287v3 (2001).

[21]     H. Sun, D. Roundy, H. J. Choi, S. G. Louie, and M. L. Cohen, to be published.

[22]     J. Kortus, I. I. Mazin, K. D. Belashchenko, V. P. Antropov, and L. L. Boyer, Phys. Rev. Lett. **86** (2001) 4656.

[23]     P. B. Allen and R. C. Dynes, Phys. Rev. B **12** (1975) 905.

[24]     H. Padamasee, J. E. Neighbor, and C. A. Schiffman, J. Low Temp. Phys. **12** (1973) 387.

[25]     F. Bouquet, Y. Wang, R. A. Fisher, D. G. Hinks, J. G. Jorgensen, A. Junod, and N. E. Phillips, to be published.

[26]     B. Mhhlshlegel, Z. Phys. **155** (1959) 313.

[27]     G. Karapetrov, M. Iavarone, W. K. Kwok, G. W. Crabtree, and D. G.Hinks, Phys. Rev. Lett. **86** (2001) 4374.

[28]     A. Sharoni, I. Felner, and O. Millo, Phys. Rev. Rev. B **63** (2001) 220508R.

[29]     C.–T. Chen, P. Seneor, N.–C. Yeh, R. P. Vasquez, C. U. Jung, M.–S. Park, H.–J. Km, W. N. Kang, and S.–I. Lee, cond–mat/0104285 (2001).

[30]     G. Rubio–Bollinger, H. Suderow, and S. Vieira, Phys. Rev. Lett **86** (2001) 5582.

[31]     F. Giubileo, D. Roditchev, W. Sacks, R. Lamy, and J. Klein, cond–mat/0105146 (2001).

[32]     F. Giubileo, D. Roditchev, W. Sacks, R. Lamy, D. X. Thanh, and J. Klein, cond–mat/0105592 (2001).

[33]     H. Schmidt, J. F. Zasadzinski, K. E. Gray, and D. G. Hinks, Phys. Rev. B **63** (2001) 220504R.

[34]     P. Szabo, P. Samuely, J. Kacmarcik, Th. Klein, J. Marcus, D. Fruchart, S. Miraglia, C. Marcenat, and A. G. M. Jansen, cond–mat/0105598 (2001).

[35]     X. K. Chen, M. J. Konstantinović, J. C. Irwin, D. D. Lawrie, and J. P. Franck, cond–mat/0104005 (2001).

[36]     T. Takahashi, T. Sato, S. Souma, T. Muranaka, and J. Akimitsu, Phys. Rev. Lett. **86** (2001) 4915.

[37]     S. Tsuda, T. Yokoya, T. Kiss, Y. Takano, K. Togano, H. Ihara, and S. Shin, cond–mat/0104489 (2001).

[38]     R. K. Kremer, B. J. Gibson, and K. Ahn, cond–mat/0102432 (2001).   (Two versions of this report have appeared under the same number.)

[39]     Ch. Wälti, E. Felder, C. Degen, G. Wigger, R. Monnier, B. Delley, and H. R. Ott, cond–mat/0102522 (2001).

[40]     A. Junod, Y. Wang, F. Bouquet, and P. Toulemonde, in this volume.

Fig. 1. (a) $[C(H) - C(9\text{ T})]/T$. In (a) and (b) the scale has been shifted by $\gamma(9\text{ T})$ to give an approximation to $C_e(H)/T$ (see text). In (a) the dashed curve is a polynomial extrapolation of the 12–20 K, $H = 0$ data to $T = 0$; the horizontal line represents $\gamma(9\text{ T})$. In (b) the low–$T$ 5– and 7-T data are shown on an expanded scale with solid curves representing fits described in the text; the horizontal solid and dashed lines represent $\gamma(9\text{ T})$ and $\gamma_n$, respectively. In (c) the solid lines represent an entropy–conserving construction. The error bars are ± 0.1% of the total measured heat capacity.

Fig 2. $\gamma$ as a function of $H$. The solid curve is a guide to the eye and an extrapolation to $H_{c2}(0)$. The dashed curve is a "fit" with an $H_{c2}$ anisotropy of 10.

Fig. 3. (a) Entropies as functions of $T$ for different $H$, with $H$ increasing from the lowest to highest curve. (b) Thermodynamic critical field, compared with a BCS curve for the derived values $\gamma_n$ and $T_c$.

Fig. 4. The low–temperature exponential dependence of $C_{es}$ on $T^1$.

Fig. 5. Fit to $C_{es}$ with a two–gap model (see text).

Figure 1

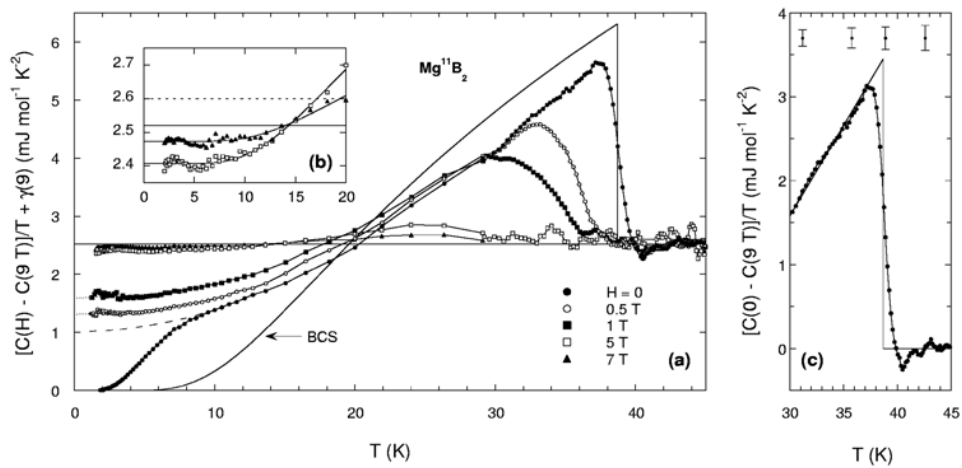

Figure 2

Figure 3

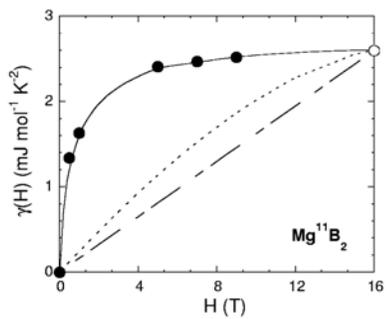

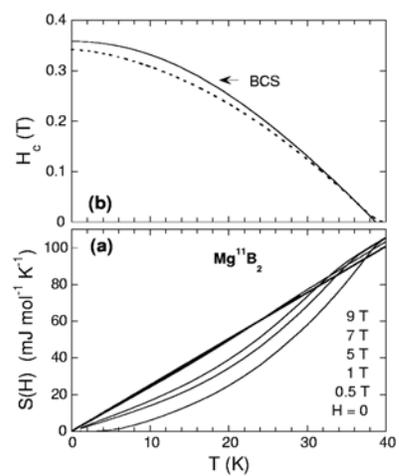

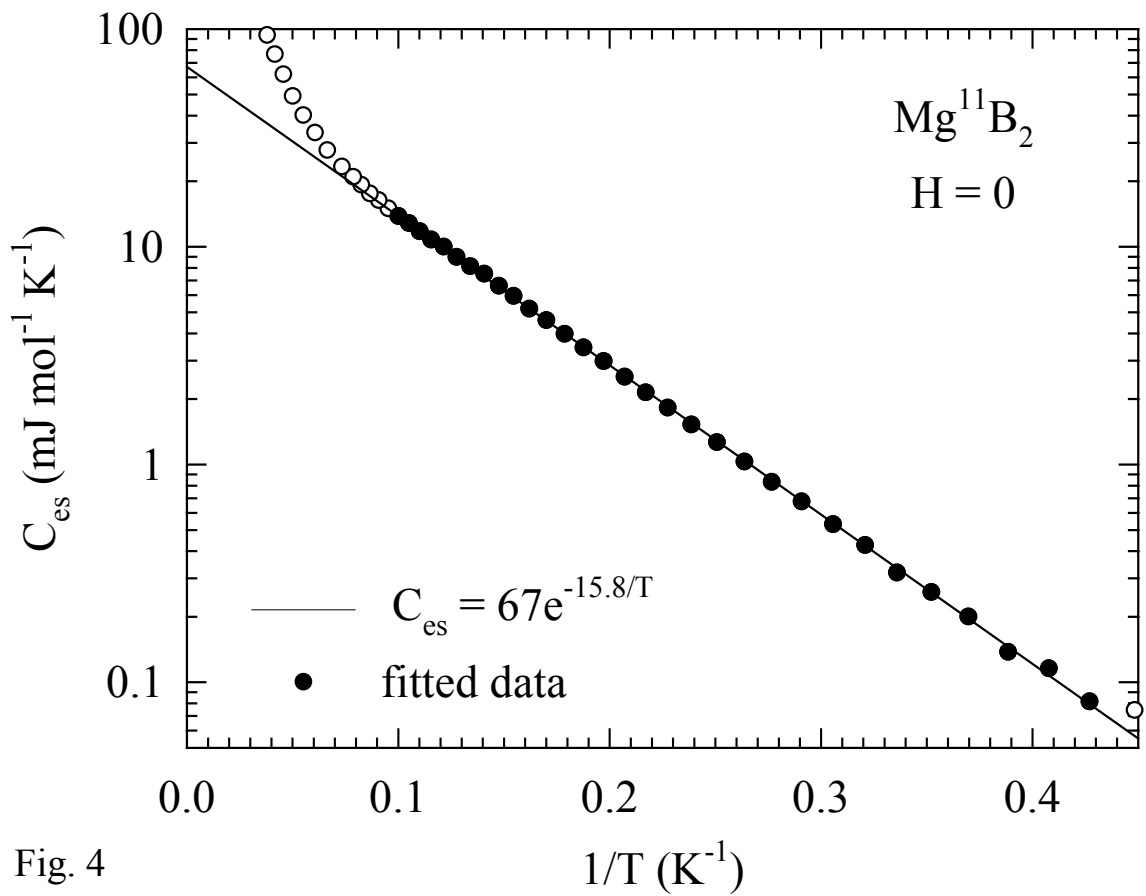

Fig. 4

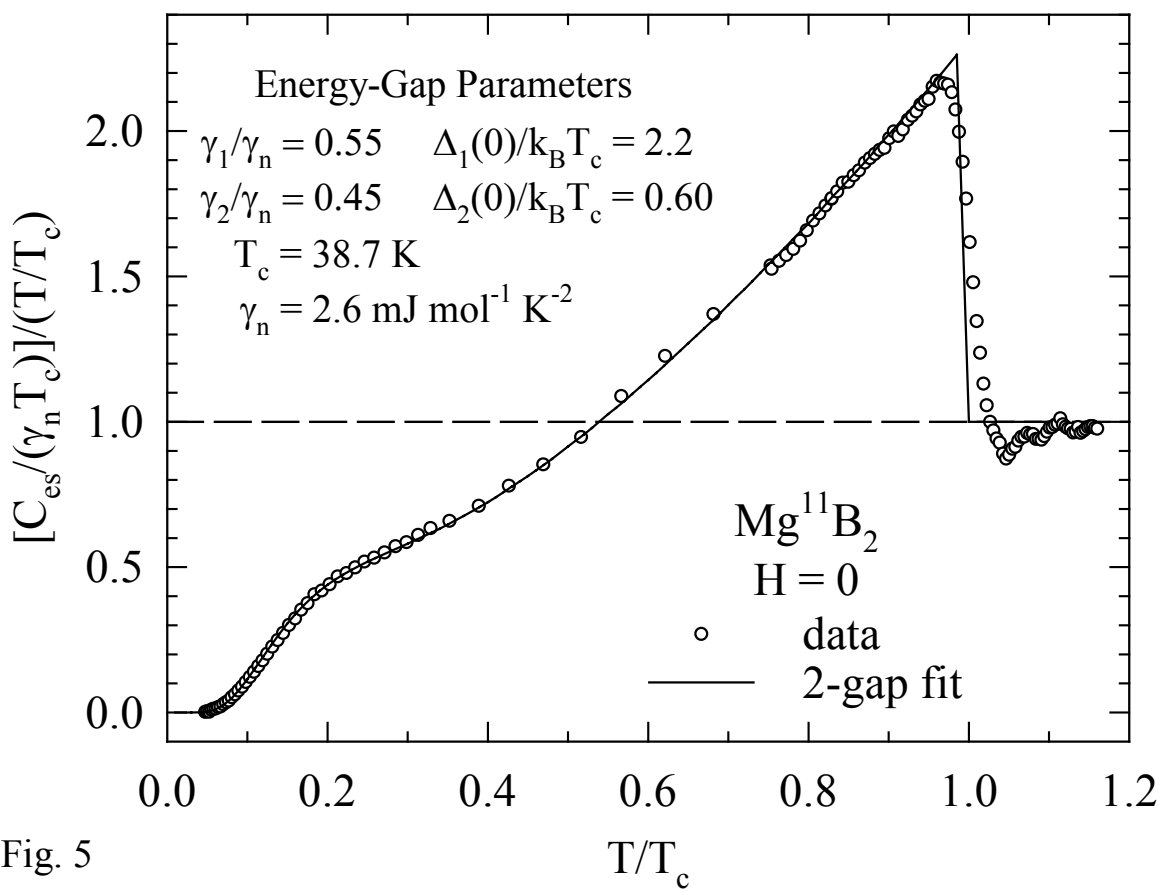

Fig. 5